\documentclass[aps,prl,reprint,twocolumn]{revtex4-1}
\usepackage{graphicx}
\usepackage{amsmath}
\usepackage{amssymb}
\usepackage{comment}
\usepackage[colorlinks, allcolors=blue]{hyperref}
\usepackage{hypcap}

\newcommand{\pref}[2]{\hyperref[#1]{\ref{#1}#2}}
\newcommand{\eqpref}[1]{\hyperref[#1]{(\ref{#1})}}

\begin{document}
\title{Direct observation of chiral currents and magnetic reflection in atomic flux lattices}
\author{Fangzhao Alex An}
\author{Eric J. Meier}
\author{Bryce Gadway}
\email{bgadway@illinois.edu}
\affiliation{Department of Physics, University of Illinois at Urbana-Champaign, Urbana, IL 61801-3080, USA}
\date{\today}
\begin{abstract}
The prospect of studying topological matter with the precision and control of atomic physics has driven the development of many techniques for engineering artificial magnetic fields and spin-orbit interactions. Recently, the idea of introducing nontrivial topology through the use of internal (or external) atomic states as effective ``synthetic dimensions'' has garnered attraction for its versatility and possible immunity from heating.
Here, we directly engineer tunable artificial gauge fields through the local control of tunneling phases in an effectively two-dimensional manifold of discrete atomic momentum states.
We demonstrate the ability to create homogeneous gauge fields of arbitrary value, directly imaging the site-resolved dynamics of induced chiral currents.
We furthermore engineer the first inhomogeneous artificial gauge fields for cold atoms, observing the magnetic reflection of atoms incident upon a step-like variation of an artificial vector potential. These results open up new possibilities for the study of topological phases and localization phenomena in atomic gases.
\end{abstract}
\maketitle

The unique experimental capabilities associated with ultracold atomic matter have made it an ideal candidate platform for the study of topological phenomena~\cite{Goldman-NatPhysReview}.
Given the purity and microscopic understanding of atomic gases, they can be used to gain insight into the nature of correlated topological states.
Additionally, the high level of control over atomic systems has enabled the exploration of topological phenomena not readily accessible in real materials.
The past decade has seen steady progress towards the realization of stable, low temperature atomic samples with nontrivial topology.
Lattice-based techniques utilizing lattice modulation~\cite{Gemelke-Shake,Struck-Shake,Jotzu-Haldane} and laser addressing~\cite{Jaksch-RAT,Aidel-Harper,Miyake-Harper} have proven capable of reaching the regimes of large effective magnetic fields and strong spin-orbit coupling, a feat that has eluded bulk gas techniques like rotation~\cite{Fetter-RMP} and bulk Raman addressing~\cite{Higbie-Raman,Spielman-Raman,Dalibard-RMP-Gauge}. Still, nontrivial heating remains an issue for lattice-based schemes~\cite{Rosch-HeatingFloquet,Mueller-Floquet}.

\begin{figure}[t!]
	\includegraphics[width=0.93\columnwidth]{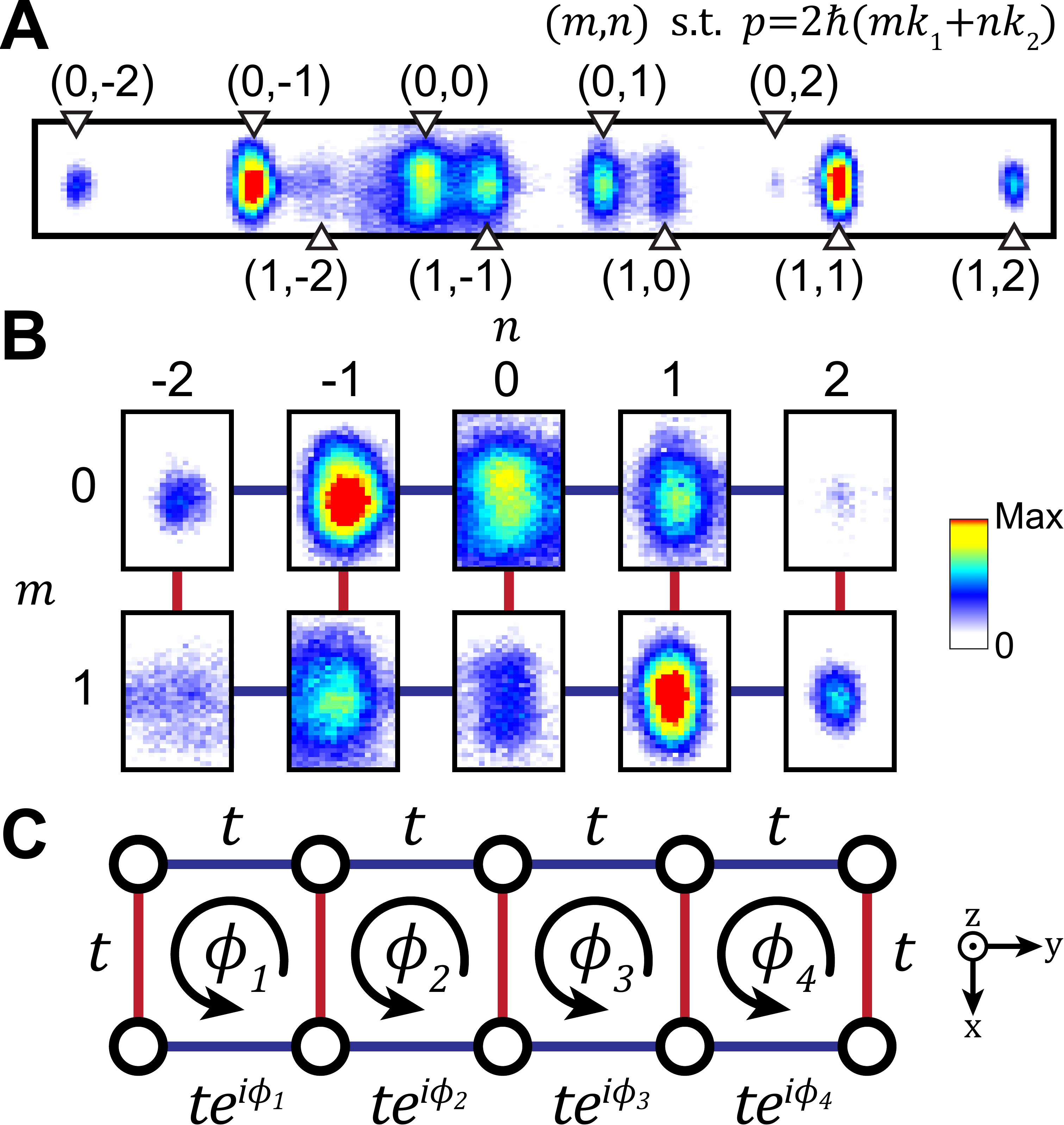}
	\caption{\label{FIG:fig1}
		\textbf{Two-leg flux ladder.}
		\textbf{(A)}~Time-of-flight image of atoms in momentum orders labeled by integer indices $(m,n)$ with momenta $p=2\hbar (m k_1 + n k_2)$.
		\textbf{(B)}~Image from (A) rearranged to show the 2D lattice. Red vertical and blue horizontal links are controlled by the $\lambda_1$ and $\lambda_2$ wavelength lattices ($k_{1(2)}=2\pi/\lambda_{1(2)}$), respectively. This figure and (A) show absorption images using the normalized optical density scale at right.
		\textbf{(C)}~Schematic of a two-leg ladder with applied tunneling phases $\phi_i$ on each link of the $m=1$ leg, resulting in fluxes $\phi_i$ around each four-site plaquette.
	}
\end{figure}

Recently, the use of atomic internal states as \emph{synthetic dimensions}~\cite{Boada-Synthetic,Celi-synthetic,Stuhl2015,Mancini2015,Wall-synthetic} has emerged as an interesting alternative strategy that may obviate some sources of heating.
While various analogues of real-space transport have previously been studied using internal~\cite{Chang-spinor-josephson} and discrete momentum~\cite{Moore-Anderson,Chabe-Anderson-2008,Gadway-rotor} states of cold atoms,
the application of spectroscopically-controlled, field-driven transitions to the study of topological matter has led to recent key developments, including the realization of two-dimensional (2D) systems with fixed artificial flux~\cite{Stuhl2015,Mancini2015}.

\begin{figure*}[t!]
	\includegraphics[width=\textwidth]{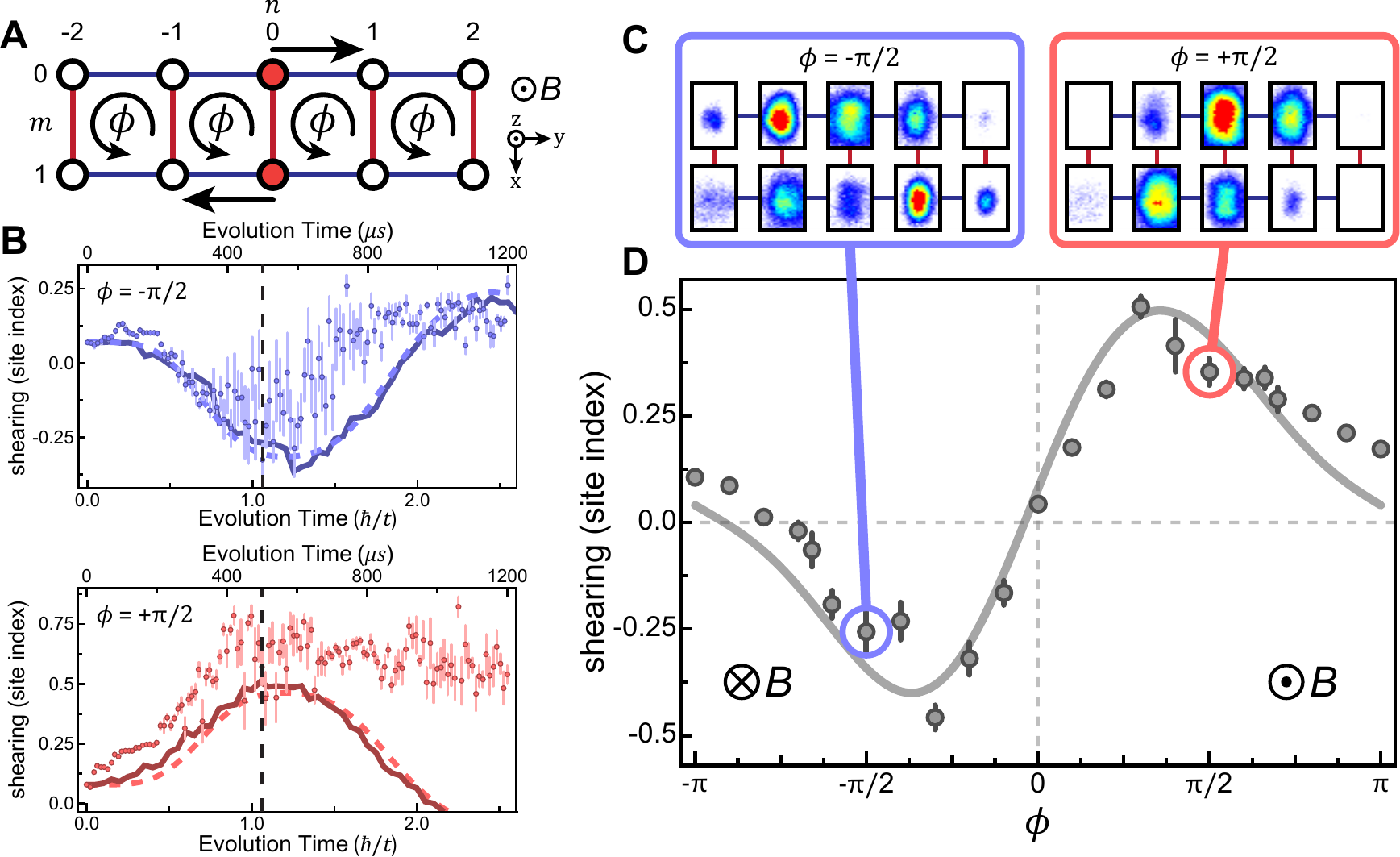}
	\caption{\label{FIG:fig2}
		\textbf{Shearing in the flux ladder.}
		\textbf{(A)}~Schematic showing atoms undergoing clockwise shear (arrows) for positive flux $\phi$, corresponding to an effective magnetic field $B$ directed out of the page. Red filled-in circles represent the initial state.
		\textbf{(B)}~Shearing dynamics for $\phi = -\pi/2$ (top, blue) and $\phi = +\pi/2$ (bottom, red). Dashed and solid curves represent numerical simulation results based on Eq.~\ref{EQ:e0} and a more complete model taking into account off-resonant transitions, respectively, both scaled and offset to match the data. Dashed vertical lines indicate the time when the data for (C) and (D) were taken.
		\textbf{(C)}~Site populations for $\phi = -\pi/2$ (left, blue) and $\phi = +\pi/2$ (right, red). Color scale used is the same as in Fig.~\pref{FIG:fig1}{B}.
		\textbf{(D)}~Shearing vs. applied flux. Solid line represents results from a simulation of the more complete model. Measurements for (C) and (D) were taken after $500$ $\mu$s ($\sim 1.06$ $\hbar/t$), indicated by dashed vertical lines in (B). The calibrated tunneling rates for (B) and (D) are slightly different, so this time translates into different tunneling times for the two. All error bars denote one standard error.
	}
\end{figure*}

Here, we expand the capabilities of synthetic dimension-based simulation by engineering fully-tunable flux lattices in \emph{multiple} synthetic dimensions.
We directly image chiral atomic currents induced by a homogeneous flux, and observe magnetic reflection of atoms from a step-like jump of an effective magnetic vector potential generated by an inhomogeneous flux. These advances in the creation of artificial gauge fields, combined with the available control of all tunneling terms and site energies, should greatly expand the variety of topological systems open to investigation through cold atom simulation.

Our implementation~\cite{Gadway-KSPACE,Meier-AtomOptics,Meier-SSH} laser couples the discrete momentum states of ultracold $^{87}$Rb atoms to mimic tunnel-coupled lattice sites.
In one dimension (1D), we drive the two-photon Bragg transitions coupling these momentum states using counter-propagating laser fields with a wavelength of $\lambda_2=1064 $ nm (wavenumber $k_2 = 2\pi / \lambda_2$).
Here we extend this scheme to higher dimensions by adding a second set of Bragg laser beams, co-propagating and having an incommensurate wavelength ($\lambda_1=781.5 $ nm, $k_1 = 2\pi / \lambda_1$) with respect to the $\lambda_2$ laser.
The Bragg laser wavevectors $k_{1,2}$ define an effective 2D manifold of discrete momentum states carrying momenta $p_{m,n} = 2\hbar(mk_1+nk_2)$. Starting with a Bose--Einstein condensate at rest, we populate these states by applying $m$ and $n$ two-photon Bragg transitions from the $k_1$ and $k_2$ lasers, respectively.
This mapping between the atoms' 1D momentum distribution and the 2D lattice with site indices $(m,n)$ is depicted in Fig.~1,~A~and~B. By imprinting a multi-frequency spectrum onto each pair of lasers, we are able to individually address every allowed transition in this 2D system with spectroscopic precision (Fig.~\pref{FIG:fig1}{C}), allowing for full control of all tunneling terms and site energies in a synthetic 2D tight-binding model~\cite{SuppMat}.

We begin by directly mimicking a magnetic vector potential in the Landau gauge, $\mathbf{\hat{A}} = (0, B x, 0)$, through coordination of the tunneling phases on a 2$\times$5-site ladder. This gives rise to a uniform effective magnetic field as shown in Fig.~\pref{FIG:fig2}{A}. The dynamics of our cold atoms are effectively governed by the Hamiltonian
\begin{equation}
\hat{H} = -[t_x \sum_{n} \hat{c}^\dag_{1,n} \hat{c}_{0,n} + t_y \sum_{m,n} e^{i \phi_{m,n}} \hat{c}^\dag_{m,n+1} \hat{c}_{m,n} ] + \mathrm{h.c.} ,
\label{EQ:e0}
\end{equation}
where $\hat{c}_{m,n}$ ($\hat{c}^\dagger_{m,n}$) is the bosonic annihilation (creation) operator for the state with indices $(m,n)$. In terms of the effective magnetic field $B$, the engineered tunneling phases along $y$ are given by $\phi_{m,n} = -m \phi$, where $\phi = 2 \pi d^2 B (q / hc)$ is the flux associated with closed loops around individual four-site plaquettes, $d$ is the effective spacing between synthetic lattice sites, $q$ is the effective charge of the particles, $h$ is Planck's constant, and $c$ is the speed of light. Here, and in the remainder of this work, we employ homogeneous tunneling strengths and engineer hard-wall system boundaries through the direct control of all tunneling magnitudes.

\begin{figure*}[t!]
	\includegraphics[width=\textwidth]{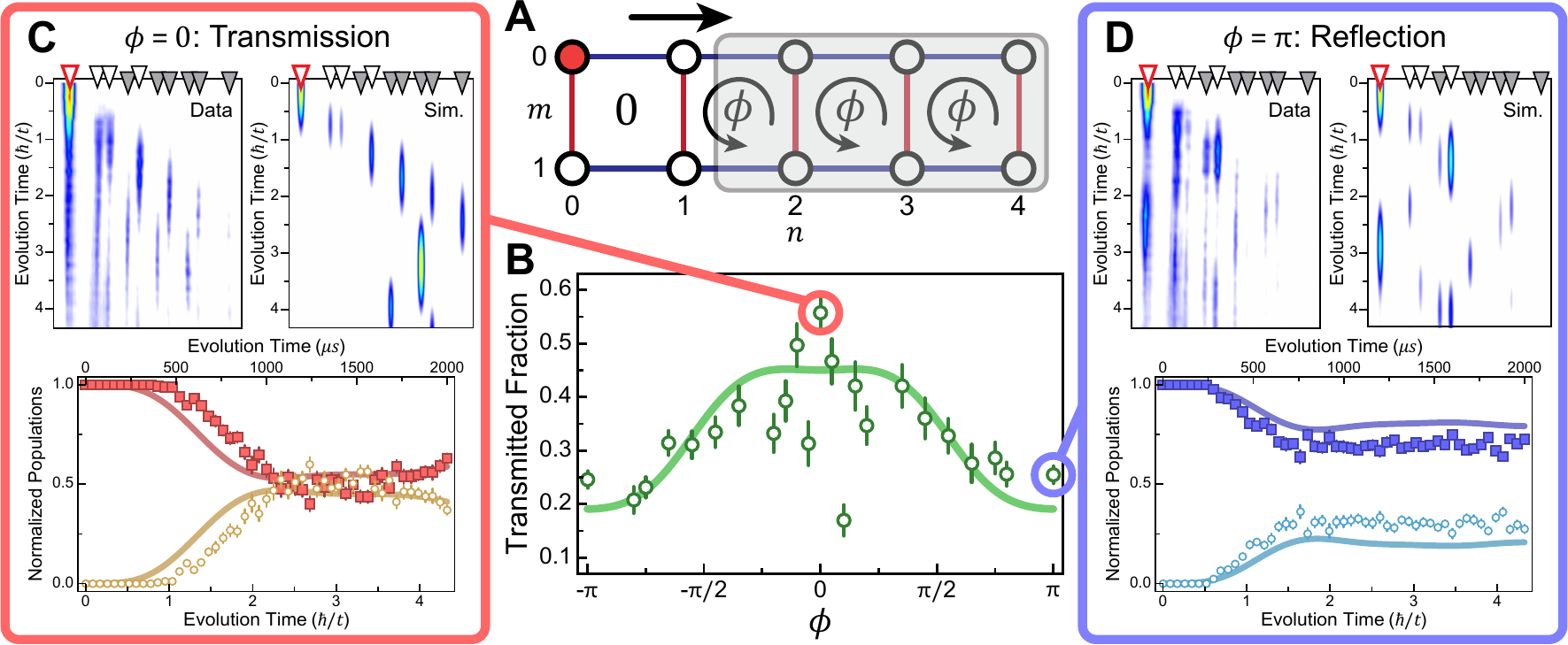}
	\caption{\label{FIG:fig3}
		\textbf{Magnetic reflection.}
		\textbf{(A)}~Schematic depicting the lattice divided into two regions of different flux: $0$ (unshaded, left) and $\phi$ (shaded, right). Population begins at the red filled-in lattice site.
		\textbf{(B)}~Fraction of initial population transmitted into the shaded $\phi$ flux region as a function of $\phi$ after $1500$~$\mu$s evolution time ($\sim 2.94$~$\hbar/t$). Solid curve represents a numerical simulation with an overall scaling factor of $0.48$  to fit the data.
		\textbf{(C} and \textbf{D)}~Dynamics for (C) $\phi=0$ and (D) $\phi=\pi$. Top: Integrated (over the image dimension normal to the lattice) optical density images vs. evolution time for data (left) and simulation (right). Left-most red marker denotes initial site and shaded gray markers denote shaded $\phi$ region. Bottom: Population in the zero flux region (darker squares) and shaded $\phi$ region (open lighter circles) as a function of evolution time.
		Calibrated tunneling time ($\hbar/t = 462(28)$~$\mu$s) for these dynamics differs from that of (B), and solid simulation curves account for an identical scaling as in (B). All error bars denote one standard error.
	}
\end{figure*}

To probe the influence of our tunable field $B$ on these ``charged'' particles, we observe their nonequilibrium response to a quench of the effective field. In particular, we study the response of atoms initially prepared in a symmetric superposition of occupation on sites $(0,0)$ and $(1,0)$.
Due to the lack of interior lattice sites, this two-leg ladder geometry does not host the same bulk localization and conductance at the boundary typical of the integer quantum Hall effect.
However, as depicted in Fig.~\pref{FIG:fig2}{A}, the applied fluxes do lead to anisotropically conducting chiral currents, or a ``shearing'' of the initial symmetric state along the $m=0$ and $m=1$ legs. We define this shearing to be
\begin{equation}
\label{EQ:shear}
\text{shearing} \equiv \langle n \rangle_0 - \langle n \rangle_1,
\end{equation}
where $\langle n \rangle_{0(1)}$ is the average site index along the $m=0$ $(m=1)$ leg. In general, application of a positive flux $\phi$ will induce a clockwise chiral current and a positive shear, as shown in Fig.~\pref{FIG:fig2}{A}. A sign reversal of the flux should result in a reversal of the shearing direction, and for fluxes of zero or $\pm \pi$ we expect only symmetric spreading of the initial state along the $y$ direction.
While recent experiments~\cite{Atala2014,Mancini2015,Stuhl2015} have observed evidence for chiral currents on similar two- and three-leg flux ladders, our use of a fully synthetic lattice allows us to engineer arbitrary fluxes, and furthermore enables direct observation of all site populations and shearing dynamics at the site-resolved level.

Figure~\pref{FIG:fig2}{B} shows the observed shearing dynamics for applied fluxes $\phi = -\pi/2$ (top, blue) and $\phi=+\pi/2$ (bottom, red). Initially, all of the population resides in the middle sites, and thus should give zero shear (see the supplement~\cite{SuppMat} regarding the small initial nonzero shear). The atoms then follow the general trend described above: positive flux causes atoms to move clockwise around the ladder, and negative flux leads to motion in the opposite direction. Due to the finite system size, the value of the shearing does not continue to grow ad infinitum, but saturates and decreases as the atoms reach the ends of the ladder and move between the two legs.
Figure~\pref{FIG:fig2}{C} shows the population distributions after a quench duration of 500~$\mu$s (dashed vertical lines in Fig.~\pref{FIG:fig2}{B}) for fluxes $\phi = \pm \pi/2$.
A clear distinction between the cases of positive and negative flux can be seen at this time, corresponding to the case of near-maximum shear.
For longer times, as seen in Fig.~\pref{FIG:fig2}{B}, the data tends to deviate from the simple theory simulations.
The dashed lines are the predictions of Eq.~\ref{EQ:e0} for a tunneling rate $t / \hbar = 2 \pi \times 338$~Hz, which exceeds the experimentally calibrated tunneling rates of Fig.~\pref{FIG:fig2}{B} and Fig.~\pref{FIG:fig2}{D} by $\sim 25\%$ and $\sim 31\%$, respectively \cite{SuppMat}.
Solid curves represent a more detailed model that includes the influence of off-resonant Bragg transitions~\cite{Gadway-KSPACE,SuppMat}, but which still ignores the influences of atomic interactions, finite condensate size, and effective decoherence due to both the phase instability of the Bragg lasers and the physical separation of wavepackets with different momenta.

Figure~\pref{FIG:fig2}{D} displays the measured shearing after 500~$\mu$s for the full range of applied flux values, demonstrating our wide control over homogeneous effective fields. While for $\phi = 0$ almost no shear is measured (corresponding to symmetric spreading along $y$), maximal shearing magnitudes are observed for flux values near $\pm \pi/2$.
The data are in excellent qualitative agreement with the theory curve, which has been scaled by a factor of $0.45$ to account for reductions of shearing due to decoherence and other influences. The majority of deviations from the idealized dynamics of Eq.~\ref{EQ:e0}, including the small, non-zero shear for zero flux, are reproduced by this theory accounting for residual off-resonant Bragg couplings~\cite{Gadway-KSPACE,SuppMat}. Our complete control of flux values is a necessary step towards measurement of the Hofstadter spectrum in cold atoms~\cite{Jaksch-RAT}.

As a second study, we for the first time engineer inhomogeneous artificial gauge fields for cold atoms, studying the transport of atomic wavepackets incident upon a sharp dislocation of the effective magnetic field.
As shown in Fig.~\pref{FIG:fig3}{A}, we engineer a step-like jump of the magnetic vector potential $\mathbf{\hat{A}}$ by fixing the flux in the left-most plaquette to zero while retaining a tunable homogeneous flux $\phi$ in the remaining plaquettes.
Without any initialization procedure, we begin with all of the population in the corner of the flux-free region on the zero momentum site $(0,0)$. By switching our couplings along $y$ to the range $n=0$ to $n=4$, we shift the lattice such that atoms with zero momentum naturally start on the corner site.
We then quench on tunneling and the full flux distribution and track the dynamics of the atomic distributions, monitoring the percentage of atoms that transmit through the step-like flux boundary, escaping the left-most four-site plaquette.

We probe the full range of $\phi$, as shown in Fig.~\pref{FIG:fig3}{B}, directly measuring the transmitted fraction of atoms after an evolution time of $1500$~$\mu$s ($\sim 2.94$ $\hbar/t$).
The tunneling rate $t/\hbar = 2\pi \times 311(14)$~Hz has been determined by calibrations to 2-site Rabi oscillations.
A clear trend is observed: maximum transmission near $\phi=0$ where the step in the vector potential vanishes, and maximum reflection for flux dislocations of $\pm \pi$. This is in good qualitative agreement with the predictions of the idealized tight-binding Hamiltonian of Eq.~\ref{EQ:e0}, shown as the green solid line in Fig.~\pref{FIG:fig3}{B}.
We note that this behavior is purely due to the presence of a flux boundary in this two-dimensional system, since no corresponding reflection is observed in one-dimensional chains with a step-like variation in tunneling phase.

While the idealized predictions of Eq.~\ref{EQ:e0} expect full transmission for $\phi=0$ (and roughly 40\% for $\phi=\pm \pi$), we observe reduced dynamics in the data, which we attribute to experimental sources of decoherence and dephasing that may be ameliorated in future investigations~\cite{SuppMat}.
Moreover we find that a sizable fraction of the atoms in our initial condensate (site $(0,-2)$) does not participate in the Bragg laser-driven dynamics.
This owes to the wide momentum spread of our finite-sized condensate compared to the sharp spectral selectivity of our weak coupling fields (with tunneling time of $\hbar / t = 511(22)$~$\mu$s). To account for these deviations (detailed in the supplement \cite{SuppMat}), we scale the predicted transmission curve by a factor of $0.48$ with no extra offsets. This scaling better matches the lessened transmission near $\phi=0$, but diverges from the data for larger values of flux where atoms should reflect off the boundary, regardless of effects that hinder transmission.

We additionally investigate the full dynamics for the cases of homogeneous zero flux ($\phi=0$) and maximally inhomogeneous flux  ($\phi=\pi$), as shown in Figs.~\pref{FIG:fig3}{,~C~and~D}. In both cases, we compare the complete momentum-state distributions to those predicted by Eq.~\ref{EQ:e0}, and extract the percentages of reflected and transmitted atoms.
The calibrated tunneling rate for these data ($t/\hbar = 2\pi \times 344(21)$~Hz) differs from the varying flux data discussed above.
The normalized integrated optical density (OD) plots for the $\phi = 0$ case in Fig.~\pref{FIG:fig3}{C} show a significant percentage of the population leaving the four left-most sites (denoted by white markers) and entering the right-most sites (shaded gray markers).
The number of transmitted atoms at times exceeds the number that remain in the four left-most sites, as shown in the reflected and transmitted population dynamics at bottom. These data agree quite well qualitatively with the theory predictions (with the same scaling as in Fig.~\pref{FIG:fig3}{B}).

The observations of significant transmission for $\phi = 0$ are contrasted by our measurements for $\phi=\pi$, shown in Fig.~\pref{FIG:fig3}{D}. Here, in the upper OD plots, good qualitative agreement is found between the measured population dynamics and the unscaled theory predictions, with population first leaving and then returning to the initial site (left-most red marker). While the $\phi=0$ case showed limited transmission, here the populations clearly display reflection from the boundary. At the bottom, we see that the number of atoms staying in the four left-most sites always significantly exceeds the number of transmitted atoms.
The theory curves have been scaled down to correct for the limited transmission near $\phi=0$, so in this case of maximal reflection, the scaling causes an underestimate of the transmitted fraction.
This observation of reflection from a flux boundary, absent any variation in the underlying potential energy landscape, is a purely quantum mechanical effect, in analogy to previous observations of quantum reflection~\cite{Pasquini-QuantumRef}.

Our capabilities to directly engineer artificial homogeneous and inhomogeneous gauge fields and to directly image site populations in a synthetic lattice are extremely promising for future realizations of myriad model systems relevant to topology and transport. These include 2D models of localization at topological interfaces~\cite{Hsieh-TopProx}, in disordered quantum Hall systems, and in random gauge fields~\cite{Lee-Fisher-RandomFlux-1981}.
While our results are predominantly driven by single-particle physics, the condensate atoms in our momentum-space lattice have a very long-ranged (nearly all-to-all) interaction energy, allowing for a straightforward extension to studies of interacting topological fluids.
This could be accomplished through either Feshbach-enhanced scattering properties, longer interrogation and coherence times, or by mapping to other forms of discrete motional eigenstates (trapped states instead of plane-wave momentum states) with a more local interaction~\cite{NateGold-TrapShake} or internal spin states~\cite{Boada-Synthetic,Celi-synthetic}.

We thank S. Hegde, K. Padavi\'{c}, I. Mondragon-Shem, S. Vishveshwara, and T. L. Hughes for helpful discussions, and J. O. Ang'ong'a for careful reading of the manuscript. During the preparation of this manuscript, we became aware of two related works that have demonstrated spin-orbit coupling using transitions to long-lived excited states in optical lattice clocks~\cite{Ye-SpinOrb,Fallani-SpinOrb}, based on the synthetic dimensions scheme suggested in Ref.~\cite{Wall-synthetic}. In particular, using this technique Ref.~\cite{Fallani-SpinOrb} has demonstrated a wide control of homogeneous artificial flux magnitudes in synthetic two-leg ladders.

\bibliographystyle{apsrev4-1}

\end{document}


\title{Supplementary Materials for ``Direct observation of chiral currents and magnetic reflection in atomic flux lattices''}
\author{Fangzhao Alex An}
\author{Eric J. Meier}
\author{Bryce Gadway}
\email{bgadway@illinois.edu}
\affiliation{Department of Physics, University of Illinois at Urbana-Champaign, Urbana, IL 61801-3080, USA}
\date{\today}
\maketitle

\renewcommand\thefigure{S\arabic{figure}}

\begin{flushleft}
\underline{Experimental Setup}
\end{flushleft}

Our experiment begins with the preparation of a $^{87}$Rb Bose-Einstein condensate with $\sim 5 \times 10^4$ atoms.
We reach quantum degeneracy via all-optical evaporation in a trap comprised of three optical dipole beams: two with wavelength $1064$ nm and one with wavelength $1070$ nm.
Immediately following evaporation, the condensate is transferred to a trap formed mainly by one of these beams (wavelength $\lambda_2 = 1064$ nm).
We then turn on a second beam (wavelength $\lambda_1 = 781.5$ nm) co-propagating with this $1064$ nm beam.
At the same time, we switch on the acousto-optic modulators (AOMs) in our lattice setup, which allow both the $\lambda_1$ and $\lambda_2$ beams to be retroreflected back towards the atoms, forming optical lattices.

\begin{figure}[t!]
	\includegraphics[width=\columnwidth]{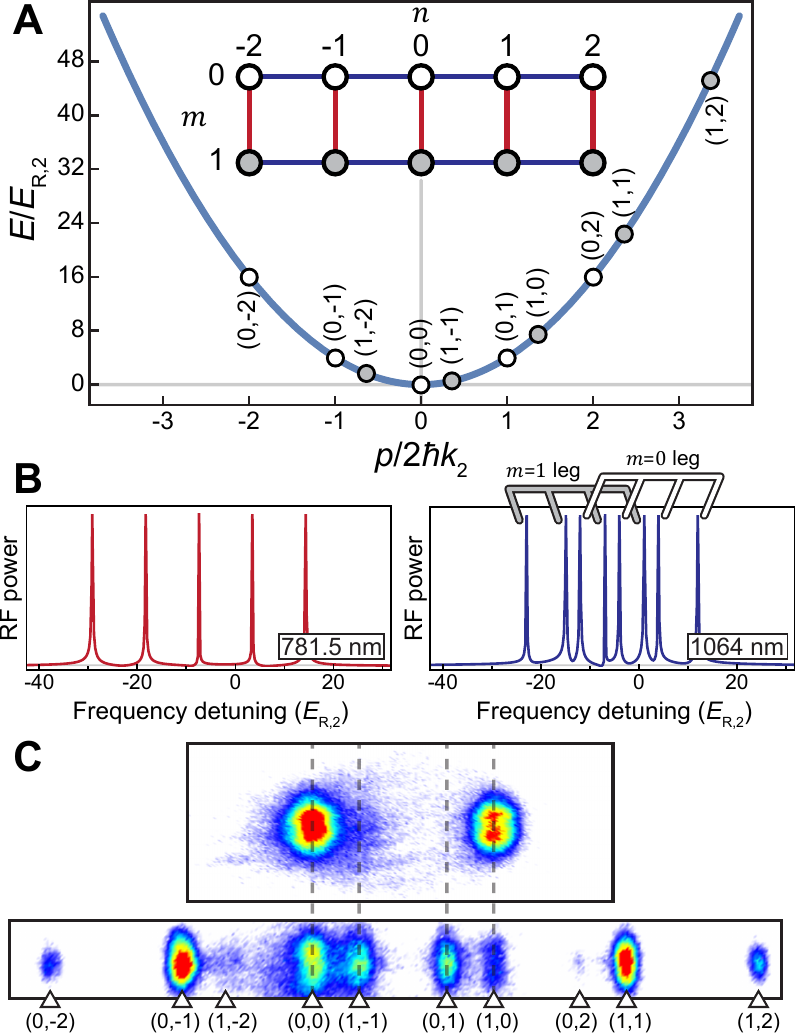}
	\caption{\label{FIG:figS1}
		\textbf{2D lattice implementation.}
		\textbf{(A)}~Free particle dispersion relation showing momentum states on the $m=0$ (white circles) and $m=1$ (gray circles) legs, labeled by $(m,n)$ with momenta $p=2\hbar(mk_1+nk_2)$. Inset: 2D lattice representation, with links addressed by the $\lambda_1$ (red, vertical) and $\lambda_2$ (blue, horizontal) wavelength beams. The recoil energy is given by $E_{R,2} = \hbar^2k_2^2/2M_{\mathrm{Rb}}$.
		\textbf{(B)}~Fourier spectra showing frequencies that address inter-leg transitions (red, left) and intra-leg transitions (blue, right). Transitions along $m=0$ and $m=1$ legs are labeled with white and gray markers, respectively.
		\textbf{(C)}~Time-of-flight images of shearing experiment after an evolution time of $0$ $\mu s$ (top) and $500$ $\mu s$ (bottom) showing nonzero initial shearing.
	}
\end{figure}

The lattice setup, described in detail in~\cite{Gadway-KSPACE,Meier-AtomOptics}, uses AOMs to write specific frequencies onto the retro beam.
These frequency components of the retroreflected beam are detuned from the single-frequency forward-propagating beam so as to resonantly address momentum-changing Bragg transitions of the atoms. 
In this work, to extend our previously introduced scheme~\cite{Meier-AtomOptics} to two synthetic dimensions, we add an identical lattice AOM setup to control the additional lattice beam (wavelength $\lambda_1$).
Figure~\pref{FIG:figS1}{A} shows where the states of the effective two-leg ladder lie on the atoms' free-particle dispersion relation, and Fig.~\pref{FIG:figS1}{B} shows the frequency teeth that we apply to the $\lambda_1$ (red, left) and $\lambda_2$ (blue, right) beams to control transitions between the two legs and along the two legs, respectively. By controlling the phases and amplitudes of these frequency components and their detunings from Bragg resonances, we can engineer arbitrary tunneling phases, tunneling amplitudes, and site energies. In this work, we address only first-order Bragg transitions between nearest-neighbor states that differ by two photon momenta, though we also have control over next-nearest-neighbor couplings relating to longer-range tunneling.

The lasers address the atoms for some duration, during which a set of Bragg transitions are driven in a phase-, frequency-, and amplitude-controlled fashion. After this evolution time, during which the dynamics of an effective tight-binding Hamiltonian are realized~\cite{Gadway-KSPACE,Meier-AtomOptics}, all of the traps are turned off and the atoms fall for 18~ms of time-of-flight (TOF). The different momentum states of our condensate atoms separate during TOF, transverse to the direction of gravity, and the populations of the various orders are measured by absorption imaging.

\begin{flushleft}
\underline{Nonzero initial shearing}
\end{flushleft}

The plots of shearing dynamics in Fig.~2B of the main text exhibit sizeable, nonzero shearing at zero evolution time. This is likely non-physical, since at zero time the lasers coupling sites along each leg (in the $n$ direction) have not been turned on.
Instead, this measured shear is likely the result of interaction effects of the atoms that occur as they fall during TOF.
The TOF image at zero time is shown in Fig.~\pref{FIG:figS1}{C} (top), along with a TOF image from the main text taken at later evolution time (bottom).
After initialization, the atoms are in a superposition of the two $n = 0$ sites on each leg of the ladder, populating only the modes $(0,0)$ and $(1,0)$. Taking a TOF image at this point causes the two orders of the condensate to separate according to their momenta.
Initially overlapped in space, the two orders fly through each other such that $s$-wave scattering leads to the pairwise redistribution of atoms into a spherical shell of momentum states about the center of mass, creating an ``$s$-wave halo''.

While this atomic ``halo'' signal is not very impressive in the 2D images shown in Fig.~\pref{FIG:figS1}{C}, it can have nontrivial consequences when we consider fits to the 1D momentum distributions of the atoms, which we obtain by integrating TOF images over the vertical direction. In particular, atomic population from this $s$-wave halo overlaps with momentum orders that should in principle not be populated, in this case the states $(1,-1)$ and $(0,1)$. As population in either of these modes contributes to a measured positive (clockwise) shear, this interaction-driven effect leads to a nonzero, positive shear signal at short times. At longer evolution times, population spreads out over more momentum orders, so that the influence of this nonlinear scattering process becomes less severe. To avoid these minor complications in future experiments, one could suppress the atomic interactions to zero during time-of-flight. In addition to atoms scattering from $s$-wave collisions, occasional atoms from the tail of the $(1,0)$ state's momentum distribution are fit as part of the nearby $(0,1)$ state, as the two orders are closely spaced with respect to the expanded size of the condensate momentum orders. The same happens to the adjacent $(0,0)$ and $(1,-1)$ orders.

\begin{flushleft}
\underline{Challenges of working in 2D}
\end{flushleft}

Here we briefly describe some technical challenges associated with the extension of our previously studied technique to two synthetic dimensions, leading to deviations between the experimental data and theory predictions at longer evolution times.
In particular, the present scheme using an effective 2D manifold of momentum states requires us to operate at a range of tunneling times that are much longer than those employed in previous 1D studies~\cite{Meier-AtomOptics,Meier-SSH}. This restriction to long tunneling times stems from the more condensed spectrum of Bragg frequencies required to generate our 2D synthetic lattice.
Figure~\pref{FIG:figS1}{A} illustrates the mapping between the momentum states and the sites of the effective 2D synthetic lattice, and Fig.~\pref{FIG:figS1}{B} shows the relevant Doppler shifts related to the Bragg resonances enabling tunneling between the legs (red) and along the legs (blue) of the ladder.
This spectrum of frequency components, each of which need to be separately spectroscopically addressed, is much denser than the 1D case.
To faithfully resolve individual resonances and to avoid off-resonant coupling to nearby transitions, we need to significantly decrease the various two-photon Rabi rates that relate to nearest-neighbor tunneling.
In practice, we are forced to reduce the tunneling rates and tunneling times by a factor of $\sim$$4$--$5$ compared to those used in our studies of 1D lattice systems~\cite{Meier-AtomOptics,Meier-SSH}.

This requirement of reduced tunneling bandwidths alone would not necessarily present any challenges, but it reduces the number of coherent tunneling events we can observe if our system features fixed timescales associated with decoherence or dephasing. Presently, one major limitation of our system is the natural loss of coherent momentum-space dynamics as the momentum states of our condensate separate and lose spatial overlap. While some effort has been put into decreasing the trapping frequencies of our optical trap along one axis to $\sim 2 \pi \times 10$~Hz, thus increasing the size of our condensate and its coherence length at low temperatures, such considerations should still presently limit us in 2D to the observation of coherent dynamics over $\lesssim 4$ $\hbar/t$. This natural source of decoherence can be ameliorated in future experiments by working with larger condensates with lower momentum-spread, or by working with trapped spatial eigenstates as suggested in Ref.~\cite{NateGold-TrapShake}.

Even ignoring this loss of near-field interference, these reduced tunneling rates still have a noticeable effect at short evolution times, as shown in Fig.~3 of the main text.
Due to the finite size of the trapped condensates, our atoms feature a spread of momenta along the direction of imparted momentum. Because the resonance frequency of each Bragg transition depends on the initial state of the atoms, this leads to a spread of relevant Bragg resonance conditions that can exceed the spectral bandwidth of the applied Bragg laser ``pulses''~\cite{Ketterle-Bragg}.
Thus, the expected dynamics of atoms in coherently-coupled momentum orders is not realized for all atoms in the finite-sized condensate, and in particular a significant portion of atoms can be seen persisting in the zeroth momentum order (site $(0,0)$) in Figs.~3, C and D, of the main text.
We expect that this source of deviation from the expected transmission and reflection dynamics for the inhomogeneous flux lattice could also be changed by working with larger, more spectrally narrow condensates.

\begin{figure}[t!]
	\includegraphics[width=\columnwidth]{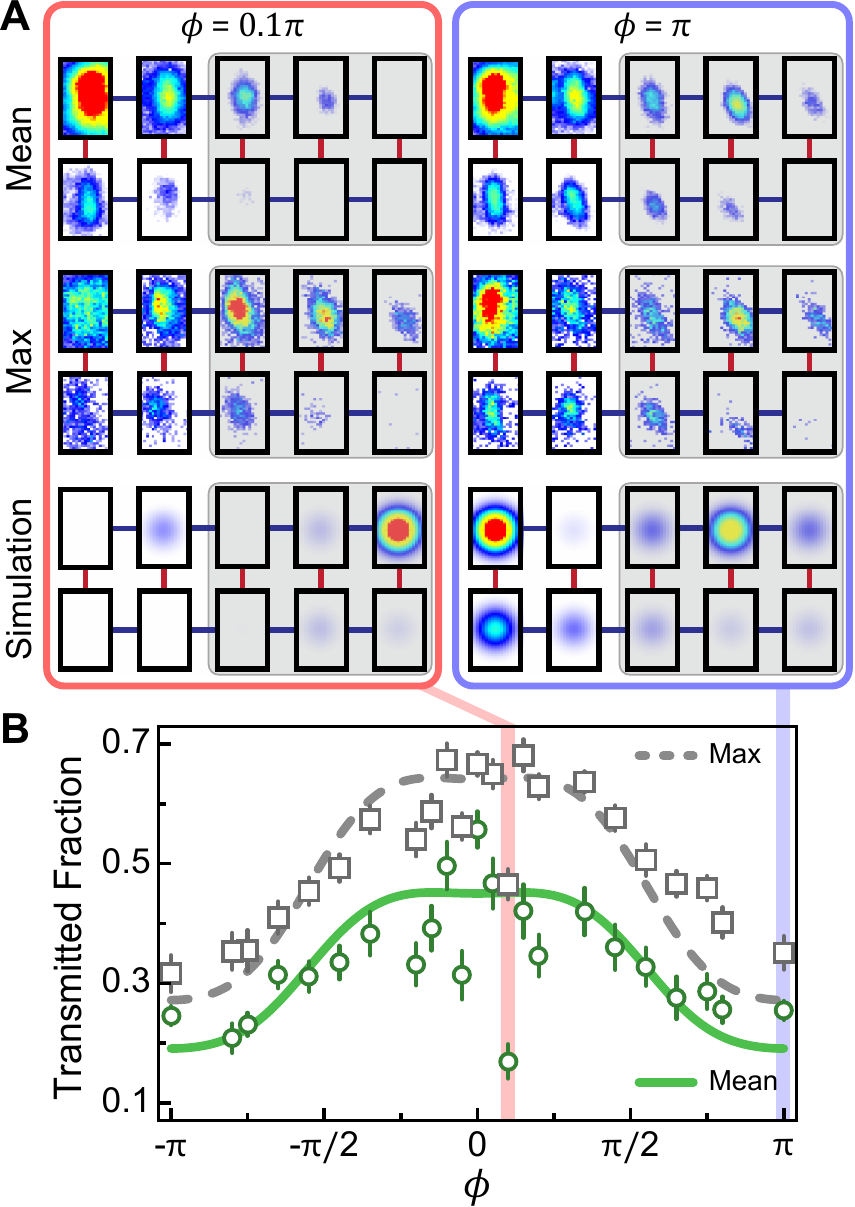}
	\caption{\label{FIG:figS2}
		\textbf{Phase instability.}
		\textbf{(A)}~Rearranged time-of-flight images of the magnetic reflection experiment for $\phi=0.1\pi$ (left) and $\phi=\pi$ (right), taken after $1500$ $\mu$s evolution time ($\sim 2.94$ $\hbar/t$). From top to bottom: average of $20$ individual images, a single image with maximum transmission, and simulated results.
		\textbf{(B)}~Transmitted fraction of atoms in the magnetic reflection experiment vs. applied flux for the average over $20$ trials (green circles, same as Fig. 3B) and the single trial with maximum transmission at each flux value (open gray squares). Simulation results have been scaled to fit the mean (solid green curve) and max (dashed gray curve) data.
	}
\end{figure}

Further, this first realization of a fully synthetic 2D lattice of states, coupled by two separate co-propagating pairs of Bragg laser fields, was achieved even without active phase stabilization of the relevant laser paths, which each contain multiple active elements (AOMs). We should thus expect to be sensitive to vibrations, thermal drifts, and other uncontrolled variations in the effective path lengths, resulting in a loss of phase coherence of the driven Bragg transitions. All of this noise is essentially common mode, i.e. shared by all relevant transitions along a given direction in the synthetic lattice, and should not affect the flux per plaquette when considered at any instance of time. However, temporal variation of the relative laser phases, which are mapped onto common-mode variations of the tunneling phase along either lattice direction, can still lead to arrested dynamics of the atomic populations. In particular, when phase drifts of order $\pi$ occur on timescales comparable to or shorter than the tunneling time, the tunneling between adjacent sites may be suppressed. More explicitly, if all tunneling phases along a particular lattice direction increase linearly with time (as may happen if a driven AOM crystal is heating up during an experiment), an effective electric field along that direction is induced~\cite{Meier-AtomOptics}. We believe that large shot-to-shot variations of the experimentally measured dynamics result from this technical phase instability, which can in future experiments be addressed by active stabilization of the lasers' relative phase~\cite{Jotzu-Haldane}.

To illustrate the magnitude of the shot-to-shot variations in the experiment, and the influence that this had on the observed departures from the expected coherent dynamics, Fig.~\pref{FIG:figS2}{A} compares images of the magnetic reflection experiment (as in Fig.~3 of the main text) for $\phi = 0.1\pi$ (left) and $\phi = \pi$ (right).
At the top are time-of-flight images averaged over $20$ experimental shots represented on the synthetic 2D lattice (Mean), the same images used to generate the transmitted fraction data as a function of $\phi$ in Fig.~3B of the main text.
In the middle we have selected the experimental run exhibiting the largest fraction of transmitted atoms (Max), which we believe in many cases may simply be the run with the least uncontrolled phase variation over the relevant time of evolution.
Both of these cases are compared to the theoretically predicted population distribution for fully coherent dynamics. In general, the Max data tends to have better agreement with the theory.
More telling is the fact that when large disagreements are seen between theory and the Mean data, a significant fraction of atoms has not left the initial condensate order.
For further comparison, Fig.~\pref{FIG:figS2}{B} shows the experimentally measured transmitted fraction, as in Fig.~3B of the main text, for both Mean (green circles) and Max (open gray squares) data. While the Mean data fluctuate somewhat wildly around $\phi=0$, the Max data show better agreement with the transmission closer to the expected 100\% and depict a more consistent trend.

Finally, we note that the aforementioned complications associated with these longer tunneling times make it difficult to directly calibrate tunneling rates via 2-mode Rabi oscillations, as discussed in \cite{Meier-AtomOptics}.

\bibliographystyle{apsrev4-1}